\documentclass[10pt,conference,table,xcdraw,screen]{IEEEtran}
\usepackage[ruled,vlined]{algorithm2e}
\usepackage{graphicx}
\usepackage{paralist}
\usepackage{tabularx}
\usepackage{balance}
\usepackage{multirow}
\usepackage{multicol}
\usepackage{pbox}
\usepackage{mathtools}
\usepackage[TABBOTCAP]{subfigure}
\usepackage{algorithmic}
\usepackage{paralist}
\usepackage{tabularx}
\usepackage{url}
\usepackage{balance}
\usepackage{multirow}
\usepackage{multicol}
\usepackage{amsmath}
\usepackage{setspace}
\usepackage{makecell}
\usepackage{caption}
\usepackage{verbatim}
\usepackage{float}
\usepackage[normalem]{ulem}
\usepackage{xspace}
\usepackage{amssymb}
\usepackage{ifthen}
\usepackage{pifont}
\usepackage{tikz}

\usepackage{hyperref}
\include{macro}

\begin{document}

\title{An Empirical Investigation into the Use of Image Captioning for Automated Software Documentation}

\author{
	\IEEEauthorblockN{Kevin Moran$^{\dagger}$, Ali Yachnes$^{*}$, George Purnell$^{*}$, Junayed Mahmud$^{\dagger}$, \\Michele Tufano$^{\ddagger}$, Carlos Bernal Cardenas$^{\ddagger}$, Denys Poshyvanyk$^{*}$, Zach H'Doubler$^{*}$}
	\IEEEauthorblockA{
\textit{$^{\dagger}$George Mason University, VA, USA}, \textit{$^{*}$William \& Mary, VA, USA}, \textit{$^{\ddagger}$Microsoft, WA, USA}\\
 \href{mailto:}{kpmoran@gmu.edu}, \href{mailto:}{ayachnes@email.wm.edu}, \href{mailto:}{gwpurnell@email.wm.edu}, \href{mailto:}{jmahmud@gmu.edu}  \\ \href{mailto:}{michele.tufano@microsoft.com}, \href{mailto:}{carlosbe@microsoft.com}, \href{mailto:}{denys@cs.wm.edu}, \href{mailto:}{pzhdoubler@email.wm.edu}\\
	}
}

\maketitle

\begin{abstract}
Existing automated techniques for software documentation typically attempt to reason between two main sources of information: code and natural language. However, this reasoning process is often complicated by the \textit{lexical gap} between more abstract natural language and more structured programming languages. One potential bridge for this gap is the Graphical User Interface (GUI), as GUIs inherently encode salient information about underlying program functionality into rich, pixel-based data representations. 
This paper offers one of the first comprehensive empirical investigations into the connection between GUIs and functional, natural language descriptions of software. First, we collect, analyze, and open source a large dataset of functional GUI descriptions consisting of 45,998 descriptions for 10,204 screenshots from popular Android applications. The descriptions were obtained from human labelers and underwent several quality control mechanisms. To gain insight into the representational potential of GUIs, we investigate the ability of four Neural Image Captioning models to predict natural language descriptions of varying granularity when provided a screenshot as input. We evaluate these models quantitatively, using common machine translation metrics, and qualitatively through a large-scale user study. Finally, we offer learned lessons and a discussion of the potential shown by multimodal models to enhance future techniques for automated software documentation.
\end{abstract}


\maketitle

\section{Introduction \& Motivation}
\label{sec:intro}

	Proper documentation is generally considered to be an integral component of building and distributing modern software systems. In fact, past studies have illustrated the general benefits of documentation during the development lifecycle~\cite{Chen:JSS'09,Kajko-Mattsson:EMSE'05,Dagenais:FSE'10,Garousi:IST'15} and the importance of technical documentation to software maintenance and evolution~\cite{Forward:DE'02}. However, despite the value of well-documented systems, modern development processes and constraints often lead to the disregard or abandonment of a range of documentation tasks~\cite{agile-manifesto,Forward:DE'02,Kajko-Mattsson:EMSE'05,Zhi:JSS'15,Fluri:WCRE'07,Chen:JSS'09}. These difficulties have given rise to a wealth of research on automated techniques that aim to ease the burden on stakeholders by generating various types of documentation for a given task. For example, existing approaches have been developed to automatically generate natural language summaries and documentation for code~\cite{Moreno:ICPC'13a,Moreno:ICPC'13,Rigby:ICSE'13,Ying:FSE'14,Ying:FSE'13,Ratol:ASE'17,Howard:MSR'13}, APIs~\cite{Treude:ICSE'16,Petrosyan:ICSE'15}, unit tests~\cite{Li:ICST'16}, bug reports~\cite{Moreno:ICSM'13,Moran:FSE'15}, release notes~\cite{Moreno:TSE'17,Moreno:FSE'14}, and commit messages~\cite{Jiang:ASE'17,Cortes:SCAM'14}, among other artifacts~\cite{Chatterjee:MSR'17,Linares-Vasquez:ISSTA'16}. 

Generally, existing techniques for automated software documentation have been concerned with modeling relationships that exist between two primary information modalities: code and natural language (NL). Unfortunately, reasoning between these two information sources is difficult due to the \textit{lexical gap} resulting from the often disparate conceptual associations that connect source code lexicon and the more abstract words and phrases used in NL descriptions~~\cite{Biggerstaff:ACM'94,Guo:ICSE'17}. Recently, this lexical gap was acknowledged as an \textit{information inference} problem in a report made by Robillard \etal~\cite{Robillard:ICSME'17}, wherein key research challenges exist in (i) inferring undocumented program properties, and (ii) discovering latent abstractions and rationales. These challenges suggest that overcoming the semantic disconnect between code and NL may require new knowledge sources that encode distinct program properties typically absent from traditional software or NL lexicon.

One source of information which has been left largely unexplored for the purposes of automated documentation is \textit{visual software data} encoded into Graphical User Interfaces (GUIs). GUI-based applications predominate modern user-facing software, as can be readily seen in the popularity of desktop and mobile apps~\cite{android-popularity}. Furthermore, high quality applications with well-designed GUIs allow \NEW{technically-inclined} users to instinctively understand underlying program features. Thus, intuitively, certain functional properties of applications are encoded into the visual, pixel-based representation of the GUI such that cognitive human processes can determine the computing tasks provided by the interface. This suggests that there are latent \textit{patterns} that exist within visual GUI data which indicate the presence of natural use cases capturing core functionality~\cite{Moran:ICPC'18}. 

Given the inherent representational power of GUIs in conveying program related information, we set forth the following hypothesis that serves as the basis for work in this paper: 

\vspace{0.5em}
\noindent \textit{The representational power of graphical user interfaces to convey program-related information can be meaningfully leveraged to support automated techniques for software documentation.} 
\vspace{0.5em}

\noindent While most existing work on automated documentation concerns itself with the dichotomy between code and NL, we posit that the latent information encoded within GUIs can aid in bridging the existing semantic documentation gap by providing an additional source of knowledge that inherently reflects program functionality. In fact, GUI-based representations of software have the potential to address the two challenges set forth by Robillard \etal~\cite{Robillard:ICSME'17}. More specifically, GUIs can aid in \textit{inferring undocumented program properties} that are inherently represented within the design of GUI controls or widgets (\eg capturing a feature which is otherwise poorly represented by low-quality code identifiers/comments). Further,  GUIs could be used as source to \textit{mine abstractions or rationales} that would otherwise remain obscure (\eg providing a use case-based explanation of a block of code connected to a GUI screen). In overcoming these challenges, we see \textit{GUI-centric documentation} having an impact on the following types of software documentation: 

\noindent\textit{\textbf{Technical Documentation:}} Developers utilize technical documentation, such as code comments or READMEs, in order to learn about the functionality and interfaces of software to support engineering tasks. Automatically generating such documentation accurately is a challenging inference problem. However, it has been shown that GUI-related code can comprise as much as half of the code in user facing programs~\cite{MemonPHD2001}. \revision{This means that graphical software data is connected in some way to large portions of GUI-based software projects \ie through GUI-event handlers, or code stipulating GUI layouts such as \texttt{\small html}.} Therefore, if automated techniques are able to effectively infer salient functionality from the GUIs, they could be combined with existing techniques and leveraged to provide automation to developers, such as comment generation or code summarization with greater feature-based context awareness. As we illustrate in this paper, GUI code/metadata appears to encode \textit{orthogonal} information compared to visual GUI data (\ie screenshots), which suggests that we may be able to infer documentation information from visual GUI data that likely can't be inferred from GUI code alone. 

\noindent\textit{\textbf{User Documentation:}} \NEW{Developers typically provide users with documentation such as tutorials or walkthroughs to help clearly illustrate software features. While some experienced users can infer functionality directly from a GUI, end-users exhibit a range of technological expertise, and many \textit{rely} upon various forms of end-user documentation~\cite{Solem:SIGDOC'85}. Thus, building techniques capable of \textit{automatically} generating such documentation would free up development effort for other critical tasks, such as bug fixing.} Beyond typical user facing software aids, GUI-centric program documentation could also enable entirely new classes of automated accessibility features, \NEW{which are sorely needed for mobile apps~\cite{Alshayban:ICSE'20}}. For example, rather than a typical text-to-speech engine, one could envision a screen-to-functionality engine that could aid a motor-impaired user with navigating the software, without extra development effort.

To investigate the potential of automated GUI-centric software documentation, we offer one of the first comprehensive empirical investigations into this new research direction's most fundamental task: \textit{\revision{generating a natural language description given a screenshot (or screen-related information) of a software GUI}}. \NEW{Given that this task underlies the various potential applications discussed above, we view this as a logical first step towards investigating the feasibility of future techniques.} To accomplish this, we collect and analyze a dataset for \textbf{C}omprehending visua\textbf{L} sem\textbf{A}ntics to p\textbf{R}edict applicat\textbf{I}on functional\textbf{TY} (the \Clarity dataset) consisting of 45,998 functional descriptions of 10,204 screenshots of popular Android apps available on Google Play. \revision{We provide a descriptive analysis of this dataset that investigates the ``naturalness'' and semantic topics of the collected descriptions by measuring cross-entropy compared to other corpora and performing a topic modeling analysis.} To learn functional descriptions of the screens from this dataset, \NEW{we customize}, train, and test four Deep Learning (DL) models for neural image captioning---three that learn from image data and one that learns from textual GUI metadata---to predict functional descriptions of software at different granularities. We evaluate the efficacy of these models both quantitatively, by measuring the widely used BLEU metric, and qualitatively through a large-scale user study. In summary, this paper's contributions are as follows:
\vspace{0.5em}
\begin{itemize}
	\item{We collect the \Clarity dataset of GUIs annotated with 45,998 functional, NL descriptions from 10,204 screenshots of popular Android apps. The NL captions were obtained from human labelers, underwent several quality control mechanisms, and contain both high- and low-level descriptions of screen functionality. While other GUI datasets exist~\cite{Deka:UIST'17,Moran:Zenodo'18}, the \Clarity dataset differs by providing an extensively labeled set of screens, akin to Flickr8K~\cite{Hodosh:JAIR'13} or MSCOCO~\cite{mscoco};}
	\item{\revision{We illustrate the underlying, natural patterns that exist in the \Clarity dataset through topic modeling.}} 
	\item{We provide an extensive quantitative and qualitative evaluation of four \NEW{tailored} DL models for image captioning using standard metrics and a large scale user study;}
	\item{We offer an online appendix with examples of model-generated descriptions and experimental data~\cite{appendix}. \revision{Our dataset, trained models, code, and evaluation scripts are open source and accessible via the appendix.}}
\end{itemize}

\section{Background}
\label{sec:background}

\subsection{The Connection between Images and NL}
\label{subsec:images-nl}

	The task of image captioning is much more difficult than that of classification or labeling, as an effective model must be able to both learn salient features from images automatically and semantically equate these features with the proper NL words and grammar that describe them. This task of semantically aligning two completely different modalities of information has led to the development of multimodal DL architectures that jointly embed NL and pixel-based information in order to predict an appropriate description of a given input image. These techniques are typically trained on large-scale datasets that contain images annotated with multiple captions, such as MSCOCO ~\cite{mscoco}, and have largely drawn inspiration from encoder-decoder neural language models traditionally applied to machine translation tasks. In this paper, we adapt three recent architectures for image captioning, \texttt{\small neuraltalk2}~\cite{Karpathy:TPAMI'17}, the \texttt{\small im2txt}~\cite{Vinyals:TPAMI'17}, and the \texttt{\small show, attend and tell (SAT)}~\cite{XuICML'15} frameworks to predict functional descriptions of software screenshots \NEW{through the use of custom pre-training and fine-tuning procedures}. Additionally, we explore the \texttt{\small seq2seq} neural language model. 

\begin{figure}[tb]
	\centering
	\vspace{-0.4cm}
	\includegraphics[width=0.9\columnwidth]{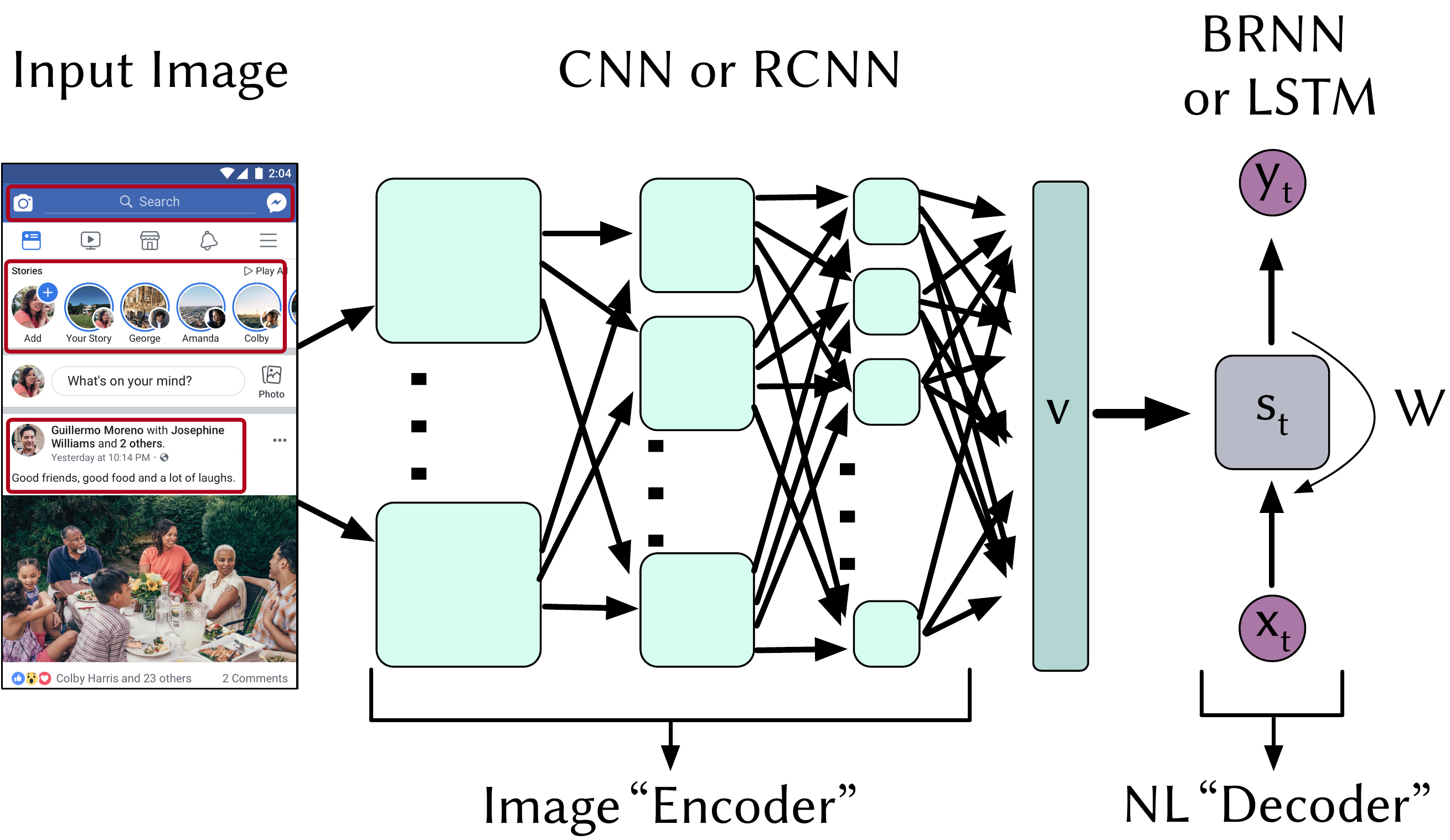}
	\vspace{-0.10cm}
	\caption{Generalized overview of multimodal DL architectures for image captioning (with RCNN)}
	\label{fig:dl-arch}
	\vspace{-0.0cm}
\end{figure}

DL models for image captioning build upon the success of encoder-decoder neural language models. The \texttt{\small im2txt} framework treats image captioning as a machine translation problem, wherein the source ``sentence'' is an image, and the target ``translation'' is a NL description. The generalized architecture of such models is shown in Fig. \ref{fig:dl-arch}. As illustrated, these architectures replace the encoder RNN with a Convolutional Neural Network (CNN), which have been shown to be highly capable of learning rich image features~\cite{Krizhevsky:NIPS12,Simonyan:ICLR'14,Szegedy:CVPR'15}. Google's implementation of \texttt{\small im2txt} uses a Long-Short Term Memory (LSTM) RNN~{\cite{Hochreiter:NC'97} for the ``decoder'' module, which has also proven extremely effective when applied to machine translation tasks. The decoder module of the \texttt{\small neuraltalk2} architecture is composed of a Bidirectional RNN (BRNN)~\cite{Schuster:TSP'97} as opposed to an LSTM. Finally, the \texttt{\small show, attend, \& tell (SAT)} model~\cite{XuICML'15} uses an LSTM decoder but with the addition of an attention mechanism that can ``attend'' to salient parts of the image representation by combining ``hard'' and ``soft'' attention mechanisms.

\vspace{-0.1cm}
\revision{
\section{Overview}
\label{sec:overview}
}
\vspace{-0.1cm}

\revision{In this section, we provide an ``at-a-glance'' overview of the data-collection procedures and various analyses performed in this paper. Figure~\ref{fig:overview} illustrates the four major components of the paper. The first major task of our study is to derive a suitable dataset of screenshot-caption pairs. We describe this process in two parts: (i) the collection of screenshots (Sec. \ref{subsec:screen-collection}), and (ii) the collection of captions from human workers (Sec. \ref{subsec:caption-collection}). The result of this data-collection effort is the \Clarity dataset, which contains 45,998 captions of 10,204 Android screenshots. Next, we aim to understand the \textit{lexical} properties of our captions through an empirical analysis in order to better understand how easily they might be modeled (Sec.~\ref{sec:data-analysis}). Thus, we perform both a comparison of the the cross-entropy of language models trained our caption corpus to other popular SE corpora, and perform an LDA-based topic analysis. Next, we discuss the process of configuring and training three neural image captioning models, and one sequence-based model to predict functional descriptions of software GUIs (Sec.~\ref{sec:modeling}). Finally, we conclude our analysis by measuring the accuracy of our trained models according to both automated reference-based metrics (\ie BLEU@$n$) and via a large-scale human evaluation. (Sec.~\ref{sec:modeling-eval})}

\begin{figure}[tb]
	\centering
	\vspace{-0.3cm}
	\includegraphics[width=\columnwidth]{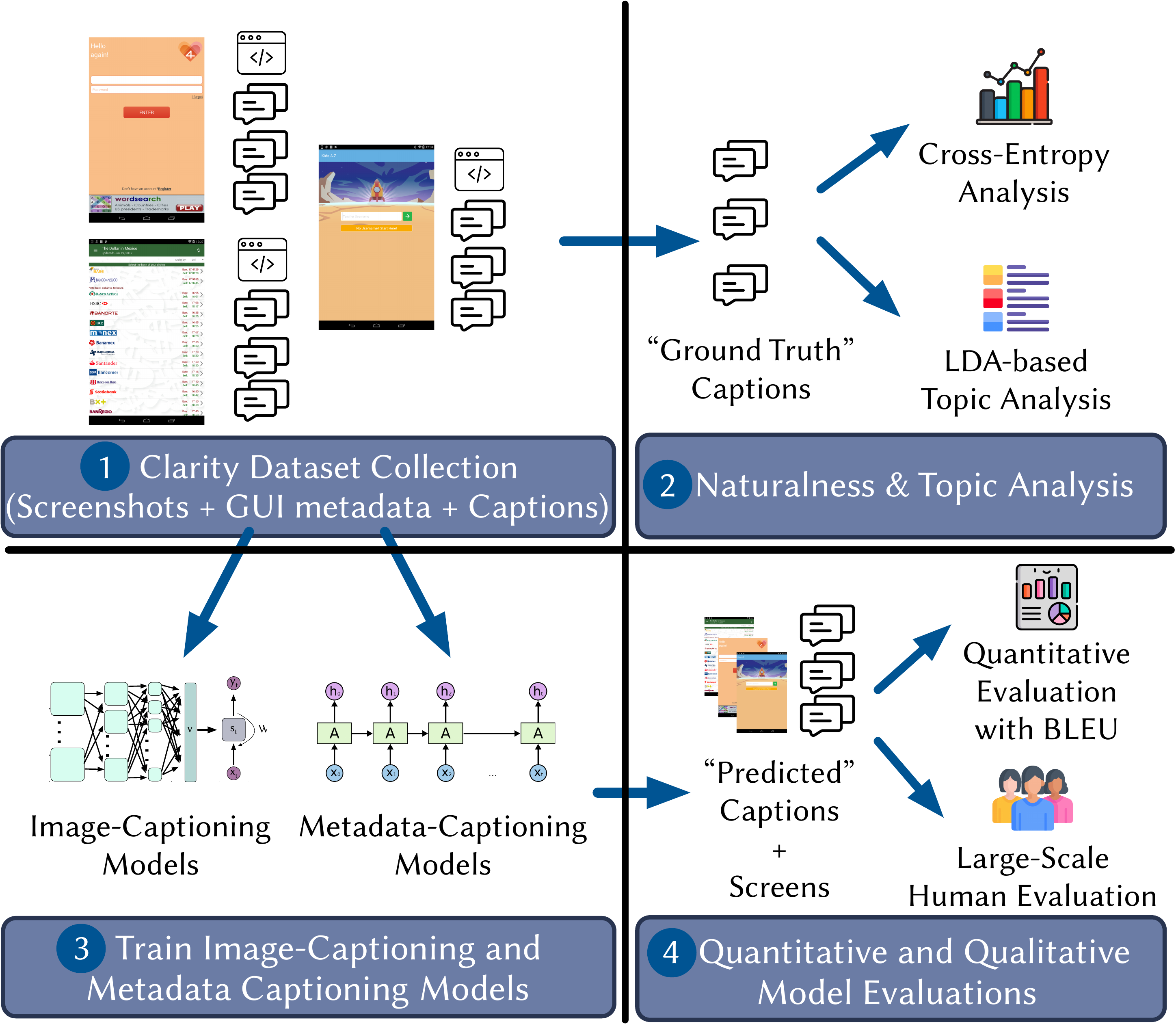}
	\vspace{-0.30cm}
	\caption{\revision{Overview of Dataset Collection and Analysis}}
	\label{fig:overview}
	\vspace{-0.0cm}
\end{figure}

\section{Dataset Collection}
\label{sec:data-collection}

\subsection{Screen \& GUI Metadata Collection}
\label{subsec:screen-collection}

	The first step in deriving the \Clarity dataset is the collection of a sizable and diverse dataset of screenshots and GUI-metadata. We chose to focus this dataset derivation on the Android platform for three main reasons: (i) Android is currently the most popular OS in the world~\cite{android-popularity}, (ii) Android apps are highly GUI-and gesture driven, making them a suitable target for our investigation, and (iii) the Android \texttt{\small screencap} and \texttt{\small uiautomator} tools facilitate the extraction of screenshots and GUI-metadata from running apps. Fortunately, large-scale datasets of Android screenshots and metadata are publicly available in related literature~\cite{Moran:TSE'18,Deka:UIST'17}. For this work, we took advantage of the \ReDraw~\cite{Moran:TSE'18,Moran:Zenodo'18} dataset which contains nearly 17k unique screenshots from 8,655 of the top-rated apps from the Google Play Store. It should be noted that another large-scale Android GUI dataset that contains a larger number of screenshots, \Rico, is also available~\cite{Deka:UIST'17}. However, we chose to utilize the \ReDraw dataset as it aligned with one of our primary study objectives. That is, we aim to learn latent feature information from both screenshots and GUI-metadata. However, for the GUI-metadata to properly align with the displayed content on a screen, the app must make use of \textit{native} Android components. Therefore, apps that primarily display their information using web technologies, so-called \textit{hybrid apps}, would obscure the GUI-metadata and impact our study findings. The \ReDraw dataset contains a set of screenshots that underwent several stages of filtering to remove instances of hybrid apps along with other noise. Furthermore, the \ReDraw dataset contains a set of GUI-component images labeled with their corresponding types (\eg \texttt{Button}) which we utilize later in our study (Sec. \ref{sec:modeling}). The end result of this filtering process was a total set of 17,203 candidate screens for labeling. We refer readers to the \ReDraw paper for complete details of the filtering process~\cite{Moran:TSE'18}. 

\vspace{-0.1cm}
\subsection{Collection of Functional Descriptions}
\label{subsec:caption-collection}

\begin{figure}[tb]
	\centering
	\vspace{-0.3cm}
	\includegraphics[width=\columnwidth]{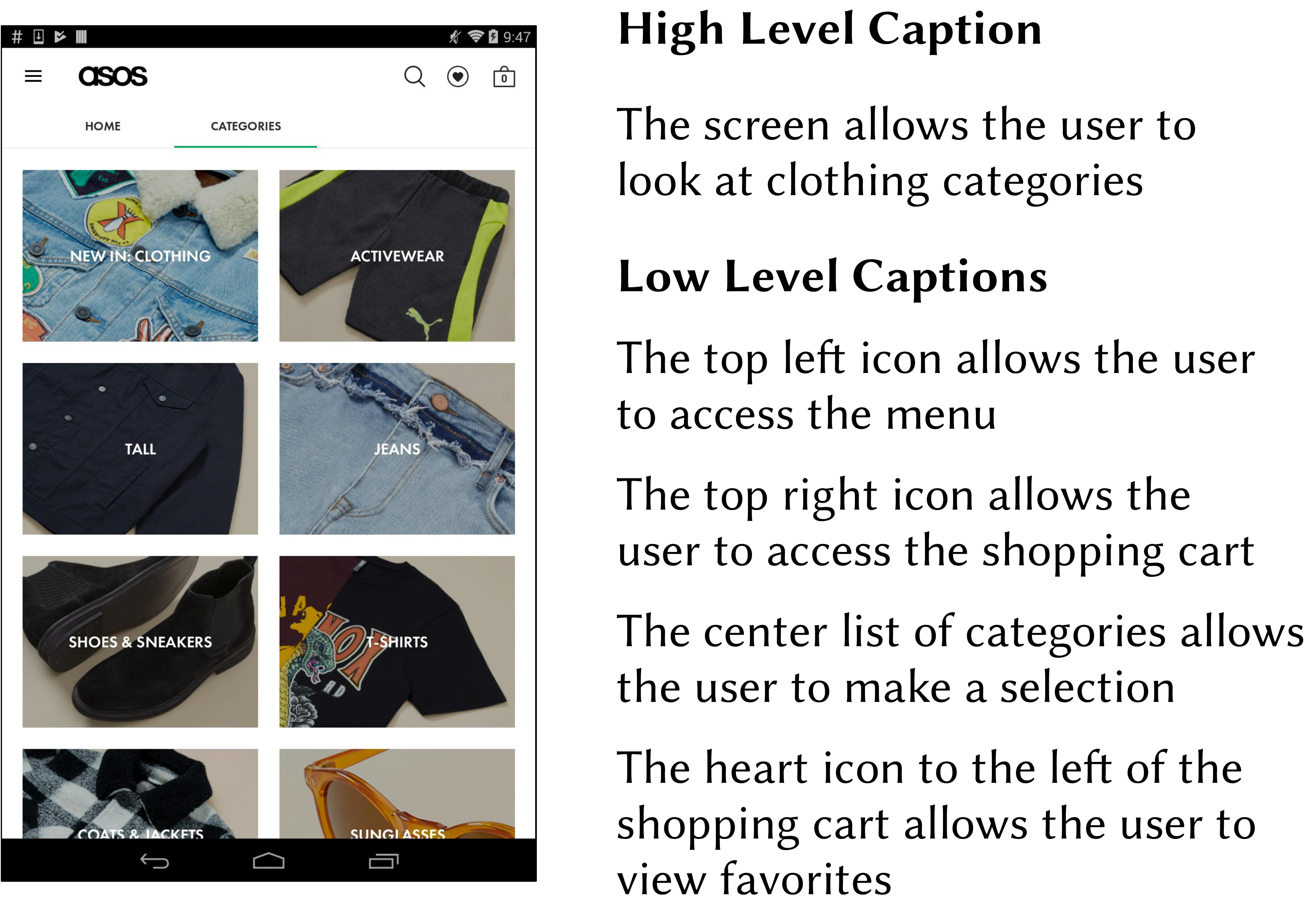}
	\vspace{-0.3cm}
	\caption{Example of captions from the \Clarity dataset.}
	\label{fig:caption-example}
	\vspace{0.0cm}
\end{figure}

	Once we derived a suitable set of screens, we needed to manually label these screens with functional captions. This process occurred in two steps: (i) first, we conducted a pilot labeling study in order to develop and prove out a tagging methodology suitable for large scale caption collection; (ii) second, we performed a full scale data collection study using Amazon's Mechanical Turk Crowd-worker platform to collect over 10k screens with functional descriptions.

\subsubsection{\textbf{Caption Granularity}} Intuitively, GUIs encode functional information at multiple levels of granularity. For example, if you were to ask a user or developer what the high-level purpose of a given screen is, they might say \textit{``This screen allows users to browse clothing categories''}, as shown in Fig. \ref{fig:caption-example}. These types of descriptions constitute the ``high-level'' functionality of a given screen. However, a single screen rarely implements only one functionality, and there may be multiple functional properties that enable the screen's high-level functional purpose. User descriptions of these types of functional properties are typically centered around the \textit{interactive} components of a screen, since these represent the instances of actions (\eg users ``doing something'') that are easily attributed to implemented functions. For example, in the screen in Fig. \ref{fig:caption-example}, underlying functions include viewing favorites, accessing a shopping cart, or selecting an item from a list. These types of ``low-level'' screen properties centered around GUI-components describe key constituent functionality. Hence, in order to capture a holistic functional view of each screen, we tasked participants with labeling each screen with one ``high-level'' functional caption, and up to four ``low-level'' functional captions. Fig. \ref{fig:caption-example} shows these two categories using actual captions collected as part of the \Clarity dataset. 

\subsubsection{\textbf{Pilot Data Collection Study}} We developed an initial image caption collection platform using a Java-based web application. Using this system, the authors manually labeled 743 screens with the caption granularities described earlier. During this study, we discovered some instances of screens with relatively little information displayed on them, making it difficult to label them with functional attributes, even after the filtering techniques discussed previously. Therefore, before moving onto the large-scale caption collection with Mechanical Turk (MTurk), at least one author manually inspected each of the 17,203 candidate screens, and discarded those with a severe lack of functionality. This resulted in a set of 16,311 candidate screens for the next phase of the study.

\subsubsection{\textbf{Mechanical Turk Data Collection Study}} To set up our large-scale data-collection process, we adapted our web application caption collection mechanism to work with MTurk's crowd worker platform. This involved configuring a Human Intelligence Task (HIT) that provided workers with a set of detailed instructions, displaying a screenshot from our dataset alongside text entry boxes for one high-level functional caption and up to four low-level functional captions (a limit was imposed to normalize the amount of time workers would spend on the HIT). This study was approved by the Institutional Review Board of the authors' affiliated institution. 

Given that we aimed to collect high-quality functional descriptions of screens in natural English, we targeted MTurk users from primarily English speaking countries that had completed at least 1,000 HITs and had a HIT approval rate of at least 90\%. We provided a detailed set of instructions for labeling images with captions that clearly explained the concept of high-level and low-level captions with examples, and provided users with explicit instructions as well as DOs and DONTs for the labeling task. The full set of instructions is available in our online appendix~\cite{appendix}.  With regard to caption quality, we specifically had three major requirements: (i) that the caption describes the perceived \textit{functionality} of a screen and not simply its appearance, (ii) that \textit{spatial references} are given for low-level captions (\eg \textit{``the button in the top-left corner of the screen''}), and (iii) that captions be written in complete English sentences with reasonably proper grammar.

	We published batches of HIT tasks by sampling unique screens from our set of 16,311 candidate screens, ensuring that no user was assigned the same screen twice. The quality of work from crowd-sourced tasks is not always optimal, so as captions were submitted, they needed to be vetted for quality. Thus, the captions for each screen were examined by at least one author for the three quality attributes mentioned above. If an author was unsure about whether a screen met these quality attributes, it was reviewed by at least one other author to reach a consensus. In total, 2,419 screens were rejected and republished as new HITs due to quality issues. In summary, 2,150 MTurk workers collected 45,998 captions (across granularities) for 10,204 screens ($\approx$5 screens per participant), and over \$2,400 was paid out.

\section{Empirical Dataset Analysis}
\label{sec:data-analysis}

	The \Clarity dataset provides a rich source of data for exploring the relationship between GUI-based and lexical software data. However, it is important to investigate the semantic makeup of the collected captions in order to better understand: (i) the \textit{latent topics} they capture as well as (ii) their \textit{naturalness} and, hence, predictability. In this section we carry out an empirical analysis of this phenomena guided by the following two Research Questions (RQs):

\begin{itemize}
	\item{\textbf{RQ$_1$:} \textit{What are the latent topics captured within the high- and low-level captions in the \Clarity dataset?}}
	\item{\textbf{RQ$_2$:} \textit{How natural (\ie predictable) are the high- and low-level captions in the \Clarity dataset?}}	
\end{itemize}

\subsection{Analysis Methodology}
\label{subsec:analysis-methods}

\begin{table}[t]
\centering
\footnotesize
\vspace{-0.2cm}
\caption{LDA topics learned over high-level captions $k=15$}
\label{tab:lda-high}
\vspace{-0.1cm}
\begin{tabular}{lllll}
\hline
Assigned Label & Top 7 Words \\ \hline \\
 "color options" & screen show app option color book differ  \\ 
 "login or create acccount" & user screen allow account log creat app \\ 
 "select image from a list" & user screen allow select view list imag   \\ 
 "map search by location" & screen locat search map user show find  \\ 
 &  &  \\ \hline
\end{tabular}
\end{table}
\begin{table}[]
\centering
\footnotesize
\vspace{-0.25cm}
\caption{LDA Topics learned on low-level captions $k=25$}
\label{tab:lda-low}
\vspace{-0.1cm}
\begin{tabular}{lllll}
\hline
Assigned Label & Top 7 Words \\ \hline \\
 "page button" & page button top center bottom side left   \\ 
 "select date" & avail date select one option theme present  \\ 
 "camera button" & video imag photo pictur  bottom camera  \\ 
 "privacy policy banner" & titl just term blue banner privaci polici  \\ 
 &  &  \\ \hline
\end{tabular}
\end{table}

\subsubsection{\textbf{RQ$_1$: Investigating Dataset Topics}} To investigate the latent topics in the \Clarity dataset, we learned topic models over caption corpora representing different granularities of functional descriptions. More specifically, we applied Latent Dirichlet Allocation (LDA)~\cite{Blei:MLR'03} to both segmented high- and low- level captions from the \Clarity dataset. In our analysis, the set of captions for a specific screenshot in the \Clarity dataset represents a document, and the entire set of captions across screenshots for a given granularity (\ie high or low level) constitutes a corpus. LDA has several configurable hyper-parameters that impact the smoothing of generated topics. These include $k$, the number of topics, $n$ which denotes the number of iterations of the sampling algorithm (Gibbs sampling~\cite{Porteous:KDD'08}, in our case), as well as $\alpha$ and $\beta$ which impact topic distributions. We set $\alpha$ and $\beta$ to standard values for NL corpora, set $n$ to 500, which proved to be a sufficient for model convergence, and varied $k$ between 15, 25, 50, and 75 topics.

\subsubsection{\textbf{RQ$_2$: Analyzing the Naturalness of GUI Descriptions}} 
Past work has pioneered the notion of the \textit{naturalness} of software~\cite{Hindle:ICSE'12}, which illustrated the fact that software, even more so than NL, exhibits repetitive patterns that make it predictable. This finding was recently further investigated and the existence of certain natural patterns was confirmed~\cite{Rahman:ICSE'19}. To illustrate naturalness, these past studies have learned statistical $n$-gram language models over software corpora, and measured the ``perplexity'' (or a log-transformed version, \textit{cross-entropy}) of these models, which represents the degree to which a model is ``surprised'' by the patterns on a test corpus when trained on a corpus from the same domain. A model with lower measured cross-entropy represents a higher predictive power, and thus, a more \textit{natural} underlying corpus.

	We follow the methodology of these past studies to explore the naturalness of the \Clarity dataset captions. Thus, similar to the methodology for the previous RQ, we split the collected captions into two corpora, one for the high-level descriptions, and one for the low-level descriptions. We then learned interpolated n-gram models, using the \texttt{\small mitlm}~\cite{mitlm} implementation of Kneser-Ney smoothing~\cite{Koehn:SMT'10}, which has been shown to be the most effective n-gram smoothing method~\cite{Hindle:ICSE'12}, following a ten-fold cross-validation procedure. We report the average cross-entropy values across these experiments for both the high and low-level corpora, compared to prior results ~\cite{Hindle:ICSE'12,Rahman:ICSE'19} for other NL and software corpora.

\subsection{Analysis Results}
\label{subsec:analysis-results}

\begin{figure}[t]
	\centering
	\vspace{-0.0cm}
	\includegraphics[width=1.1\columnwidth]{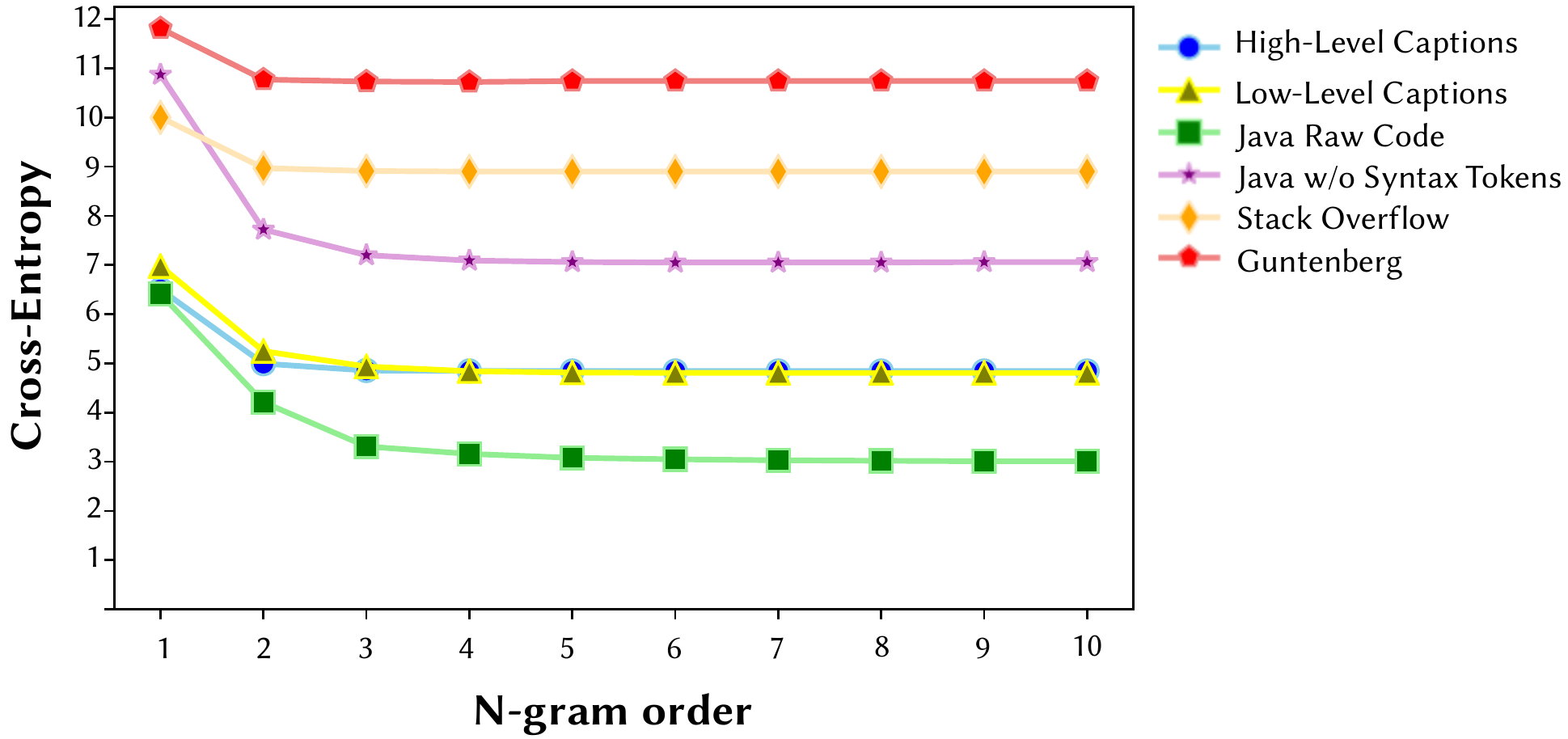}
	\vspace{-0.3cm}
	\caption{Cross-entropy of the \Clarity dataset's high and low-Level captions compared to other corpora.}
	\label{fig:cross-entropy}
	\vspace{-0.0cm}
\end{figure}

\subsubsection{\textbf{RQ$_1$: Results of Dataset Topic Modeling}}  We present selected results of some of the most representative topics in Tables \ref{tab:lda-high} \& \ref{tab:lda-low}, complete with descriptive labels that we provide for readability, and include all the results in our appendix~\cite{appendix}. These topics help to provide a descriptive illustration of some of the latent patterns that exist in both the high and low level \Clarity captions. The high-level captions illustrate several screen-level topics, including searching on a map and adjusting app settings. The low-level captions conversely capture topics that describe component-level functionality, such as date selectors, camera buttons, and back buttons. These results indicate the existence of logical topics specific to the domain of GUIs in our collected captions.

\subsubsection{\textbf{RQ$_2$: The Naturalness of Clarity Descriptions}} The results of our naturalness analysis are illustrated in Figure \ref{fig:cross-entropy}. This figure shows the average cross entropy of the high- and low- level captions from the \Clarity dataset compared to several other corpora as calculated by Rahman \etal~\cite{Rahman:ICSE'19}. More specifically, the graph depicts the average ten-fold cross entropy for: (i) The Gutenberg corpus containing over 3k English books written by over a hundred different authors, (ii) Java code from over 134 open source projects on GitHub, (iii) Java without \textit{Syntax Tokens} (\ie separators, keywords, and operators), and (iv) a Stack Overflow corpus consisting of only the English descriptions from over 200k posts.

	As described earlier, the lower the cross-entropy is for a particular dataset, the more \textit{natural} it is. That is, the corpora that exhibit lower cross entropy tend to exhibit stronger latent patterns that can be effectively modeled and predicted. As we see from Fig. \ref{fig:cross-entropy}, the \Clarity high and low level captions are \textit{more natural} than every dataset excluding raw Java code. It should be noted that, comparatively, there are several factors that could account for the observed lower cross entropy of the \Clarity captions. For instance, such factors could include other corpora \NEW{having a larger size} or having a more diverse set of human authors and writing styles. However, we mainly provide entropy measures of other datasets to provide context for the predictability of the \Clarity dataset compared to other popular corpora.  Regardless of dataset differences, the average $\approx$ 5 bits of entropy measured for the two datasets of \Clarity captions signals that our collected descriptions exhibit strong semantic patterns that can be effectively modeled for prediction. Additionally, we observe that the cross-entropy for the high and low-level captions are surprisingly similar. Intuitively, one might expect that the low-level \Clarity captions would exhibit more prevalent patterns due to the repetitive use cases of certain GUI-components such as menu buttons. This indicates the tendency of both datasets to exhibit patterns that can be appropriately modeled. However, as we illustrate in Sec. \ref{sec:modeling-eval} the ability for GUI-related information to predict captions differs according to granularity.

\section{Deep Learning Functional Descriptions from Software GUIs}
\label{sec:modeling}

	The results of the analysis from the previous section demonstrate the presence of the latent patterns in the \Clarity dataset of screenshots and captions. 
	In this section, we detail our methodology for investigating the capability of different \NEW{customized} DL models to learn these patterns to predict functional descriptions from two GUI representations.

\subsection{Clarity Dataset Segmentation}
\label{subsec:image-model-configs}

	We collected two different granularities of captions from users to derive the \Clarity dataset (Sec. \ref{subsec:caption-collection}). For the experiments in this section, we want to explore the model's ability to learn both high- and low-level functional descriptions. Thus, we split the \Clarity dataset into two groups, one containing only high level captions, and one containing only low level ones. We also created a third dataset combining the high and low level captions, in order to explore whether the predictive capabilities of the models improved by aggregating multiple granularities. It should be noted that each low-level caption was treated as a single caption (i.e. each low-level caption was treated as a separate data point) as is convention with datasets containing multiple captions~\cite{mscoco}. Each screen in the dataset has both an associated screenshot and a GUI-metadata file. In order to make for a fair comparison of performance across various model configurations, we created consistent training, validation and test partitions (80\%, 10\%, 10\% according to the number of images/GUI metadata files) to be used across models. The NL text used as input to the models was preprocessed according to the specific requirements for each model implementation~\cite{im2txt,neuraltalk2,seq2seq}.

\begin{table}[t]
\centering
\scriptsize
\vspace{-0.2cm}
\caption{Image Captioning Model Configs. used in Study}
\label{tab:image-cap-configs}
\vspace{-0.1cm}
\begin{tabular}{|l|l|l|p{2.1cm}|}
\hline
\rowcolor[HTML]{C0C0C0} 
\multicolumn{1}{|c|}{\cellcolor[HTML]{C0C0C0}\textbf{Model}} & \multicolumn{1}{c|}{\cellcolor[HTML]{C0C0C0}\textbf{Identifier}} & \multicolumn{1}{c|}{\cellcolor[HTML]{C0C0C0}\textbf{Caption Config.}} & \multicolumn{1}{p{2.7cm}|}{\cellcolor[HTML]{C0C0C0}\textbf{Model Config.}}                 \\ \hline
                                                             & im2txt-h-imgnet                                                  & High                                                                  &                                                                                     \\ \cline{2-3}
                                                             & im2txt-l-imgnet                                                  & Low                                                                   &                                                                                     \\ \cline{2-3}
                                                             & im2txt-c-imgnet                                                  & Combined                                                              & \multirow{-3}{2.7cm}{inception v3 trained on imagenet}                                  \\ \cline{2-4} 
                                                             & im2txt-h-comp                                                    & High                                                                  & \multicolumn{1}{p{2.7cm}|}{}                                                               \\ \cline{2-3}
                                                             & im2txt-l-comp                                                    & Low                                                                   & \multicolumn{1}{p{2.7cm}|}{}                                                               \\ \cline{2-3}
                                                             & im2txt-c-comp                                                    & Combined                                                              & \multicolumn{1}{p{2.7cm}|}{\multirow{-3}{2.7cm}{Inception v3 fine-tuned on Component Dataset}} \\ \cline{2-4} 
                                                             & im2txt-h-fs                                                      & High                                                                  &                                                                                     \\ \cline{2-3}
                                                             & im2txt-l-fs                                                      & Low                                                                   &                                                                                     \\ \cline{2-3}
\multirow{-9}{*}{im2txt}                                     & im2txt-c-fs                                                      & Combined                                                              & \multirow{-3}{2.7cm}{Inception v3 fine-tuned on Full Screen Dataset}                    \\ \hline
                                                             & ntk2-h-imgnet                                                    & High                                                                  &                                                                                     \\ \cline{2-3}
                                                             & ntk2-l-imgnet                                                    & Low                                                                   &                                                                                     \\ \cline{2-3}
                                                             & ntk2-c-imgnet                                                    & Combined                                                              & \multirow{-3}{2.7cm}{VGGNet pre-trained on ImageNet}                                    \\ \cline{2-4} 
                                                             & ntk2-h-ft                                                        & High                                                                  &                                                                                     \\ \cline{2-3}
                                                             & ntk2-l-ft                                                        & Low                                                                   &                                                                                     \\ \cline{2-3}
\multirow{-6}{*}{NeuralTalk 2}                               & ntk2-c-ft                                                        & Combined                                                              & \multirow{-3}{2.7cm}{VGGNet pre-trained on ImageNet with Fine Tuning}                   \\ \hline                                                             & sat-h                                                        & High                                                                  &                                                                                     \\ \cline{2-3}
                                                             & sat-l                                                        & Low                                                                   &                                                                                     \\ \cline{2-3}
\multirow{-3}{*}{SAT}                               & sat-c                                                        & Combined                                                              & \multirow{-3}{2.7cm}{VGGNet pre-trained on ImageNet}                   \\ \hline

\end{tabular}
\end{table}
\begin{table}[t]
\centering
\footnotesize
\vspace{-0.0cm}
\caption{Subset of Model Hyper-paramters}
\label{tab:hyperparameters}
\vspace{-0.1cm}
\begin{tabular}{|l|p{0.6cm}|p{1.4cm}|l|l|}
\hline
\rowcolor[HTML]{C0C0C0} 
\textbf{Hyperparameter} & \textbf{im2txt} & \textbf{NeuralTalk2} & \textbf{SAT} & \textbf{Seq2Seq}\\ \hline
Batch Size              & 64              & 16                   & 17               & 64\\ \hline
Embedding Size          & 512             & 512                  & 512              & 128\\ \hline
Decoder RNN Size/Units  & 512             & 512                  & 1024              & 128\\ \hline
Optimizer               & SGD             & SGD                 & Adam              & Adam\\ \hline
Initial Learning Rate   & 2               & 2               & 0.001                & 0.0001\\ \hline
Dropout Probability     & 0.7             & 0.7               & 0.3                 & 0.8\\ \hline
\end{tabular}
\end{table}

\subsection{Image Captioning Model Configurations}
\label{subsec:image-model-configs}

	We customize, train, and test the three neural image captioning models, \texttt{im2txt}, \texttt{neuraltalk2}, and \texttt{show, attend, \& tell (SAT)} (Sec. \ref{subsec:images-nl}), on the screenshots and captions of the \Clarity dataset. We choose to explore these three models due to their different underlying design decisions related to the type of utilized CNNs and RNNs (Sec. \ref{subsec:images-nl}), as these differences may affect their performance in our domain. \NEW{It should be noted that in the course of our experiments, we make several customizations to these models through adaptions to pre-training and fine-tuning procedures.} However, given the typical number of parameters that constitute these models, the training time can be quite prohibitive, even on modern hardware. Thus, to control our experimental complexity and investigate a number of model configurations that can be trained in a reasonable amount of time, we fix the values of the hyper-parameters for each model in our experiments. We derived our utilized hyper-parameter values by conducting random searches for optimal values of certain parameters, and chose optimal parameters reported in prior work for others. While we fix the hyper-parameters for these models, we instead customize the configurations of our image captioning models at the architectural level. Specifically, we investigate how training the ``encoder'' CNN using different datasets and training procedures effects the efficacy of the model predictions. This type of analysis allows us to more effectively flush out broader patterns related to the benefits and drawbacks of model design decisions. In the end, we trained more than 15 different configurations of the models (see Table \ref{tab:image-cap-configs}) over several machine months of computation.

\subsubsection{\textbf{im2txt Model Configurations \& Training}} For \texttt{im2txt}, we adapted Google's open source implementation of the model in TensorFlow~\cite{im2txt}. Given the incredibly large number of parameters that need to be trained for the \texttt{im2txt} model, performing even relatively simple hyperparamter searches proved to be computationally prohibitive for our experiments. Therefore, for this model we utilized the optimal set of parameters reported by Vinyals \etal~\cite{Vinyals:TPAMI'17} on similarly sized datasets. A subset of these hyper-parameter values are given in Table \ref{tab:hyperparameters}, whereas full configuration details can be found in our appendix. The publicly available implementation of Google's \texttt{im2txt} model utilizes the Inception v3~\cite{Szegedy:CVPR'16} image captioning architecture as its encoder CNN. 

In past work, the inception model weights were initialized by training on the large-scale image classification dataset ImageNet~\cite{imagenet}, which contains ``commonplace'' image categoires. However, given that we are applying these models to very particular domain (predicting descriptions of software) it is unclear if an Inception v3 model trained on the broader ImageNet dataset would capture subtle semantic patterns in the \Clarity dataset. Therefore, we explored three different model configurations to explore this phenomena: one with Inception v3 pre-trained on ImageNet, and two with Inception v3 fine-tuned on domain specific-datasets. The first domain specific image dataset we utilize is the ReDraw cropped image dataset outlined in Sec. \ref{subsec:screen-collection}, which contains over 190k images of native Android GUI-components labeled with their  type (\eg \texttt{\small Button}, \texttt{\small TextView}). The second domain specific image dataset we use consists of the full screenshots from the \Clarity dataset, labeled with their Google Play categories.

\subsubsection{\textbf{NeuralTalk2 Model Configurations \& Training}} For \texttt{\small neuraltalk2}, we adapted Karpathy \etals implementation written in Torch and lua~\cite{neuraltalk2}. We performed a brief randomized hyper-parameter search for this model, given its more efficient training time, using the optimal \texttt{\small im2txt} parameters as a starting point. The optimal values resulting from this search are provided in Table~\ref{tab:hyperparameters}. For its CNN decoder, \texttt{\small neuraltalk2} makes use of a VGGNet~\cite{Simonyan:ICLR'14} architecture pre-trained on the ImageNet~\cite{imagenet} dataset. Unlike our im2txt configurations, we explore the effect of jointly fine-tuning \texttt{\small neuraltalk2}'s CNN and RNN. Thus, we explore two configurations of \texttt{\small neuraltalk2}, one that jointly fine tunes the pre-trained VGGNet on the \Clarity dataset, and one that does not perform fine-tuning. We followed a training procedure similar to that of our \texttt{\small im2txt} models, in that we trained our models on the high, low, and combined \Clarity caption training data for 500K iterations, saving model checkpoints every 2K iterations.

\subsubsection{\textbf{Show, Attend and Tell Model Configurations \& Training}} For the \texttt{\small SAT} model, we adapted the open-source implementation of the model in Tensorflow~\cite{satCode}. The hyperparameters that we used to train our model are shown in Table~\ref{tab:hyperparameters}. The implementation used VGG16~\cite{Simonyan:ICLR'14} as its encoder CNN. We trained the \texttt{\small SAT} model on the \Clarity dataset for the low, high and combined captions for 500K iterations and kept the checkpoints after every 1K iterations. Note that due to the prohibitive training cost of this model, we did not explore using a fine-tuned VGGNet as we did with \texttt{\small neuraltalk2}.
	
\subsection{Metadata Captioning Model Configurations}
\label{subsec:image-model-configs}

	To explore the ability to translate between the lexical representations of GUI-metadata and NL functional descriptions, we train and test an encoder-decoder neural language model using Google's \texttt{seq2seq}~\cite{seq2seq} framework. Note that recent work has proposed new models that take advantage of structural text properties~\cite{Sha:AAAI'17}, however, implementations of such models are generally not available, hence we leave the study of more advanced models for future work. We chose to utilize the default general-purpose architecture and hyper-parameters for this model, as they have been shown to be effective across a wide-range of machine translation tasks~\cite{Britz:2017}. More specifically, our encoder network consists of a BRNN with Gated Recurrent Units (GRUs) and our decoder network consists of an RNN with LSTM units; hyperparameters are listed in Table~\ref{tab:hyperparameters}. 

\begin{table}[]
\centering
\scriptsize
\vspace{-0.2cm}
\caption{Metadata Captioning Model Congfigurations}
\label{tab:seq2seq-config}
\vspace{-0.2cm}
\begin{tabular}{|l|l|l|p{2.5cm}|}
\hline
\rowcolor[HTML]{C0C0C0} 
\multicolumn{1}{|c|}{\cellcolor[HTML]{C0C0C0}\textbf{Model}} & \multicolumn{1}{c|}{\cellcolor[HTML]{C0C0C0}\textbf{Identifier}} & \multicolumn{1}{c|}{\cellcolor[HTML]{C0C0C0}\textbf{Caption Config.}} & \multicolumn{1}{p{2.5cm}|}{\cellcolor[HTML]{C0C0C0}\textbf{Model Config.}}  \\ \hline
                                                             & seq2seq-h-type                                                   & High                                                                  &                                                                      \\ \cline{2-3}
                                                             & seq2seq-l-type                                                   & Low                                                                   &                                                                      \\ \cline{2-3}
                                                             & seq2seq-c-type                                                   & Combined                                                              & \multirow{-3}{2.5cm}{Trained on GUI Component Types}                     \\ \cline{2-4} 
                                                             & seq2seq-h-text                                                   & High                                                                  & \multicolumn{1}{p{2.5cm}|}{}                                                \\ \cline{2-3}
                                                             & seq2seq-l-text                                                   & Low                                                                   & \multicolumn{1}{p{2.5cm}|}{}                                                \\ \cline{2-3}
                                                             & seq2seq-c-text                                                   & Combined                                                              & \multicolumn{1}{p{2.5cm}|}{\multirow{-3}{2.5cm}{Trained on GUI-Component Text}} \\ \cline{2-4} 
                                                             & seq2seq-h-tt                                                     & High                                                                  &                                                                      \\ \cline{2-3}
                                                             & seq2seq-l-tt                                                     & Low                                                                   &                                                                      \\ \cline{2-3}
                                                             & seq2seq-c-tt                                                     & Combined                                                              & \multirow{-3}{2.5cm}{Trained on GUI-Component Type + Text}               \\ \cline{2-4} 
                                                             & seq2seq-h-ttl                                                    & High                                                                  &                                                                      \\ \cline{2-3}
                                                             & seq2seq-l-ttl                                                    & Low                                                                   &                                                                      \\ \cline{2-3}
\multirow{-12}{*}{Seq2Seq}                                   & seq2seq-c-ttl                                                    & Combined                                                              & \multirow{-3}{2.5cm}{Trained on GUI-component Type + Text + Location}    \\ \hline
\end{tabular}
\end{table}

To investigate the representative power of different attributes included in Android GUI-metadata, we create four configurations of GUI-metadata consisting of different attribute combinations (Table~\ref{tab:seq2seq-config}). We chose to utilize these attribute combinations as they represent (i) the attributes that are most likely to have values, and (ii) represent a wide range of information types (\eg displayed text, component types, and spatial information). Note that \texttt{seq2seq} did not consistently converge for the high level caption dataset, thus we do not report these results. 
Consistent with the training of the other models, our implementation of the \texttt{seq2seq} model was trained to 500k iterations, with checkpoints every 2k iterations.

\section{Deep Learning Model Evaluation}
\label{sec:modeling-eval}

	To explore our core hypothesis set forth at the beginning of this paper, and evaluate our DL models described in Sec. \ref{sec:modeling}, we perform a comprehensive empirical evaluation with two main \textit{goals}: (i) intrinsically evaluate the predictive power of the models according to a well accepted machine translation effectiveness metric, and (ii) extrinsically evaluate the models by examining and rating the quality of the predicated functional NL descriptions. The \textit{quality focus} of this evaluation is our studied models' ability to effectively predict accurate, concise, and complete functional descriptions. To aid in achieving our study goals, we define the following RQs:

\begin{itemize}
	\item{\textbf{RQ$_3$:} \textit{How accurate are our model's predicted NL descriptions?}}
	\item{\textbf{RQ$_4$:} \textit{How accurate, complete, \& understandable are our model's predicted NL descriptions from the viewpoint of evaluators?}}
\end{itemize}

\subsection{Evaluation Methodology}
\label{subsec:eval-methodology}

\subsubsection{\textbf{RQ$_3$: Empirically Evaluating Model Accuracy}} To evaluate the accuracy of our trained model's generated captions, we follow past work~\cite{Karpathy:TPAMI'17,Vinyals:TPAMI'17} and report BLEU scores~\cite{Papineni:ACL'02} of the predicted captions on the shared \Clarity test set of images and GUI-metadata. The BLEU score is a standard metric used in machine translation research that measures the textual similarity between a predicted caption (the output from a model) and a reference caption (the collected descriptions from humans in the \Clarity test set). The BLEU score can be measured according to the similarity of different subsequence lengths (\ie BLEU$_n$), and we report BLEU$_1$ through BLEU$_4$, as well as a composite score calculated as the average of these, as is convention~\cite{Karpathy:TPAMI'17,Vinyals:TPAMI'17}. For the image captioning models, we use the \textit{coco-caption} implementation of the BLEU score adapted for the \Clarity test set. For each test image across all image captioning models, three captions were generated using a beam width of 3 for the beam search across candidate predictions. The seq2seq models were evaluated in the same manner. We chose to utilize a beam width of 3 as an initial qualitative examination of our models' predictions showed this size to achieve a reasonable balance between prediction accuracy and model confidence. For the high-level captions, the three candidate captions were compared to the reference, and the overall average BLEU$_n$ scores were calculated for each model. For the low-level and combined captions, the predicted captions and reference captions were compared in a pairwise manner and overall average BLEU$_n$ scores were calculated for each model configuration.

\begin{table}[t]
\centering
\scriptsize
\vspace{-0.3cm}
\caption{BLEU Score Evaluation Results for Models}
\label{tab:bleu}
\vspace{-0.2cm}
\begin{tabular}{|l|l|l|l|l|l|l|l|}
\hline
\rowcolor[HTML]{C0C0C0} 
\textbf{Model}                & \textbf{Capt.}                 & \textbf{Model Type}                   & \textbf{B$_c$}            & \textbf{B$_1$}                  & \textbf{B$_2$}                  & \textbf{B$_3$}                  & \textbf{B$_4$}                  \\ \hline
                             & High             & im2txt-h-fs         & 12.4                         & 24.8                         & 12.6                         & 6.7                          & 5.3                          \\ \cline{2-8} 
                              & Low              & im2txt-l-comp       & 27.0                         & 45.6                         & 31.8                         & 20.0                         & 10.1                         \\ \cline{2-8} 

\multirow{-3}{*}{\texttt{im2txt}}      & Comb.         & im2txt-c-comp       & 30.3 & 51.7 & 35.9 & 22.1 & 11.6 \\ \hline
                              & High             & ntk2-h-imgnet       & 13.3 & 27.4 & 13.5 & 7.3  & 5.3  \\ \cline{2-8} 
                              & Low              & ntk2-l-ft           & 27.4 & 47.5 & 32.8 & 19.5 & 9.6  \\ \cline{2-8} 
                              
\multirow{-3}{*}{NeuralTalk2} & Comb.         & ntk2-c-ft           & 30.1                         & 52.1                         & 36.0                         & 21.8                         & 10.8                         \\ \hline
                              & Low              & seq2seq-l-type      & 18.1                         & 44.6                         & 17.0                         & 7.9                          & 0.24                         \\ \cline{2-8} 
\multirow{-2}{*}{\texttt{seq2seq}}     & Comb.         & seq2seq-c-type      & 16.9                         & 38.9                         & 14.7                         & 6.0                          & 0.08                         \\ \hline

& High             &   sat-h     & \cellcolor[HTML]{DAE8FC}17.7 & \cellcolor[HTML]{DAE8FC}30.1 & \cellcolor[HTML]{DAE8FC}18.3 & \cellcolor[HTML]{DAE8FC}12.9  & \cellcolor[HTML]{DAE8FC}9.8  \\ \cline{2-8} 
                              & Low              &   sat-l         & \cellcolor[HTML]{DAE8FC}35.0 & \cellcolor[HTML]{DAE8FC}52.5 & \cellcolor[HTML]{DAE8FC}38.7 & \cellcolor[HTML]{DAE8FC}28.1 & \cellcolor[HTML]{DAE8FC}20.7  \\ \cline{2-8} 
\multirow{-3}{*}{\texttt{SAT}} & Comb.         &     sat-c      & \cellcolor[HTML]{DAE8FC}37.7                         & \cellcolor[HTML]{DAE8FC}56.8                         & \cellcolor[HTML]{DAE8FC}42.0                         & \cellcolor[HTML]{DAE8FC}30.5                        & \cellcolor[HTML]{DAE8FC}22.0                         \\ \Xhline{2\arrayrulewidth}

NeuralTalk2             & \multicolumn{2}{l|}{Trained on Flickr8K}         & 34.0                         & 57.9                         & 38.3                         & 24.5                         & 16.0                         \\ \hline   
NeuralTalk2                    & \multicolumn{2}{l|}{\multirow{3}{*}{Trained on MSCOCO}} & 40.7                         & 62.5                         & 45.0                         & 32.1                         & 23.0                         \\ \cline{1-1}\cline{4-8}  
\texttt{im2txt}             & \multicolumn{2}{l|}{}         & 42.6                         & 66.6                         & 46.1                         & 32.9                         & 24.6                         \\ \cline{1-1}\cline{4-8}
\texttt{SAT}             & \multicolumn{2}{l|}{}         & 45.7                         & 71.8                         & 50.4                         & 35.7                         & 25.0                         \\ \hline    
\end{tabular}
\vspace{0.3cm}
\end{table}

\subsubsection{\hspace{-0.1cm}\textbf{RQ$_4$: Human Perceptions of Predicted Captions}} To qualitatively evaluate our studied model's generated captions, we performed a large-scale study involving an additional 220 participants recruited from MTurk. We randomly sampled 220 screens from the \Clarity test set, and then predicted high, low, and combined captions for them using the optimal configurations of \texttt{\small im2txt}, \texttt{\small NeuralTalk2}, and \texttt{\small seq2seq} according to the composite BLEU score for each model and caption level combination. The \texttt{SAT} captions were not included in this study due to time constraints related to the model's training. We created a HIT wherein each participant viewed 11 screenshots paired with captions. Two of the 11 captions were reference high and low to serve as a control, while the other 9 captions came from the model predictions. Screens and caption pairs were arranged into HITs such that 1) no single HIT had two of the same screenshot, 2) each of the 11 types of captions (2 reference, 9 model) were included only once per HIT. The order of these captions was randomized per HIT to prevent bias introduced by identical caption ordering between HITs. By this arrangement, each screen-caption pair was evaluated by 11 participants. After viewing these screenshot-caption pairs, participants were asked to answer six evaluation questions. Three of these questions (EQ$_1$-EQ$_3$) were adapted from prior work that assessed the quality of automatically generated code summaries~\cite{Moreno:TSE'17}, and inquired about \textit{accuracy}, \textit{completeness}, and \textit{understandability}, respectively. The three remaining questions (EQ$_4$-EQ$_6$), were free response and asked participants to explain \textit{accuracies}, \textit{inaccuracies}, and \textit{improvements}. The full set of questions and HIT are in our online appendix~\cite{appendix}. Similar to the \Clarity dataset collection, each participant's response was thoroughly vetted by at least one author, and discarded if the answers were incomplete. Responses were collected until 220 HITs were completed by unique respondents. 

\subsection{Evaluation Results}
\label{subsec:eval-results}

\subsubsection{\textbf{RQ$_3$ Results: Evaluating BLEU Scores}} We illustrate the BLEU score results for the most effective model configuration and checkpoint across all of our trained models in Table \ref{tab:bleu}, whereas the results for other model configurations can be found in our online appendix~\cite{appendix} in addition to caption examples. The cells highlighted in blue illustrate the highest performing model configuration for each caption type. In general we observe that \texttt{\small SAT} exhibits the highest overall BLEU scores across all caption granularities. We speculate that this is attributable to the addition of the advanced attention mechanism in this model that is able to ``focus'' on varying image regions or features to effectively handle multiple caption granularities. In general, the \texttt{\small seq2seq} model performed quite poorly across the varying caption types, indicating a lower tendency for rich representation. Perhaps most interestingly, we see that the optimal model configurations for the \texttt{im2txt} framework were those where the CNN was conditioned on domain specific datasets. More specifically, the best high-level caption model was conditioned on full screenshots and the best low-level caption was conditioned on the cropped GUI-component screenshots. 

Another general trend that emerges is the low-level and combined caption models tend to exhibit higher overall BELU scores compared to the high-level captions. This is somewhat intuitive, as it indicates that there are more natural connections between visual GUI and lexical patterns  in the low-level captions, compared to the high-level captions that reflect more abstract functional descriptions. When examining the captions generated by the optimal configurations of each model, it is clear that \texttt{\small im2txt} and \texttt{\small SAT} produces a more diverse set of output captions than \texttt{\small neuraltalk2}, which could be considered as more useful in many software documentation tasks.

Finally, it is worth discussing how the BLEU scores of our models compare to those of the same models trained on the more traditional Flickr8k~\cite{Hodosh:JAIR'13} and MSCOCO~\cite{mscoco} datasets given at the bottom of Table~\ref{tab:bleu}. Given the data-intensive nature of our DL models, and the much larger size of the MSCOCO dataset ($\approx$123k images, each with 5 captions), we did not expect our models trained on the \Clarity dataset to outperform those trained on MSCOCO. Thus, unsurprisingly, we observe that on average, \texttt{\small im2txt}, \texttt{\small neuraltalk2}, and \texttt{\small SAT} models trained on the MSCOCO dataset outperform the same models trained on the \Clarity datasets by $\approx$ 10 BLEU score points for the combined and low level captions, and $\approx$ 27 points on high-level captions. However, when we examine the performance of Neuraltalk2 on the more similarly sized Flickr8K dataset ($\approx$ 8K images, each with 5 captions) we observe comparable performance to the \Clarity low-level and combined datasets, with the \texttt{\small SAT} model narrowly outperforming the Flickr8K \texttt{\small neuraltalk2} model, with a slightly bigger discrepancy for the high-level captions. Overall, these results indicate that when compared with datasets of similar size, DL models trained on the \Clarity dataset exhibit similar performance.

\begin{figure}[tb]
	\centering
	\vspace{-0.3cm}
	\includegraphics[width=0.95\columnwidth]{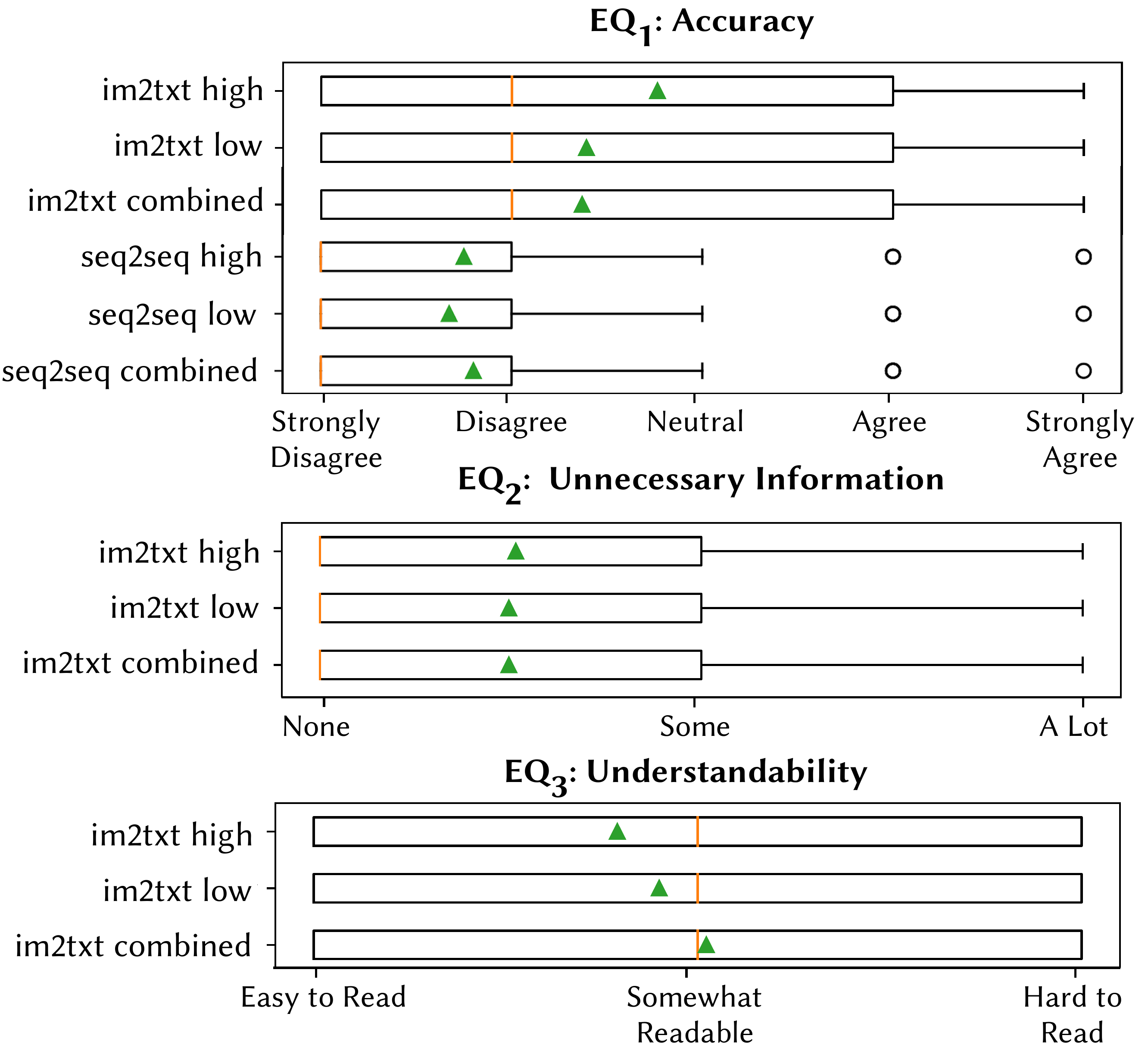}
	\vspace{-0.2cm}
	\caption{\revision{Responses across models for EQ$_1$-EQ$_3$}}
	\label{fig:user-feedback}
\end{figure} 

\subsubsection{\textbf{RQ$_4$ Results: Human Evaluations}} \NEW{The results of EQ$_1$-EQ$_3$ for the model configurations with the best performance during the human study, in addition to the \texttt{\small seq2seq} accuracy scores, are summarized in Fig.~\ref{fig:user-feedback}. Complete results across all model configurations can be found in our online appendix. The responses to EQ$_4$-EQ$_6$ varied by the type of caption, and are provided in our appendix in full. Generally, \texttt{\small im2txt} fared the best in terms of accuracy, and was followed by \texttt{\small neuraltalk2} and \texttt{\small seq2seq} respectively. For \texttt{\small im2txt}, despite mixed reactions from participants, in many cases respondents verified that the caption was accurate (\eg \textit{"The description accurately describes the screen, it is in fact a terms and conditions screen."}) and suggested minor improvements similarly to the reference captions (\eg \textit{"It could add specifics about what the settings pertain to (i.e. security)"}). As illustrated in Fig.~\ref{fig:user-feedback} the \texttt{im2txt} predictions were consistently rated as being readable and containing relevant information. It is also interesting to note that there appears to a mismatch between the performance as indicated by BLEU scores, and human perceptions, with the participants consistently rating the \texttt{im2txt} captions better than other models across EQ$_1$-EQ$_3$, despite \texttt{\small neuraltalk2} achieving a higher BLEU score for two model configurations.}

\section{Discussion \& Learned Lessons}
\label{sec:discussion}

\textbf{Lesson 1: Functional Descriptions of GUIs exhibit a high degree of naturalness and can be modeled using DL techniques.} We observed that DL models trained on the low-level and combined datasets exhibit similar performance to models trained on general image captioning datasets of similar size (\eg Flickr8K). \revision{This indicates that GUI screenshots could be used to augment approaches for automated documentation.}

\textbf{\revision{Lesson 2: GUI-centric software documentation models benefit from being pre-trained on domain specific GUI data, as opposed to general image datasets (e.g., MSCOCO)}} The qualitative results of our model analysis illustrate that for \texttt{\small im2txt}, the most effective configurations were those trained on domain specific CNN datasets. This suggests a perceptible difference between the utility of image features learned from general datasets, compared to those learned on datasets more specific to software. This suggests that future work aiming to leverage DL models for GUI-centric program documentation should look to collecting and extracting features from large-scale GUI-related datasets.

\textbf{\revision{Lesson 3: Future automated approaches for GUI-centric program documentation would likely benefit from combining the orthogonal semantics of screenshots and GUI-metadata.}} \revision{Our evaluation in this paper illustrates that the representational power of screenshots appears to be superior when applied to a software documentation task. However, given stark differences between these two modalities of information, we also observed that they encode orthogonal semantic patterns that could be combined for more effective documentation generation. One property we observed of certain captions generated by the image-based models was the effect of their limited vocabulary.  For example, certain predicted captions similar to the following: \textit{``The screen allows the user to select a \texttt{<UNK>}''}, wherein the \texttt{UNK} token represents missing token, which should be mapped to some unobserved app property, such as a ``album cover'' or ``store location''. However, such predictions could be combined with the vocabulary present in GUI metadata to help predict more complete, and accurate descriptions. Thus, a promising direction for future work is to \textit{jointly encode} both screenshots and lexical GUI-metadata.}

\textbf{\revision{Lesson 4: Training image captioning models to predict specific or diverse pieces of functionality is difficult.}} \revision{Practical models for GUI-centric documentation should able to predict both specific pieces of information (e.g. the functionality of a particular button for a given method handler), and diverse functionality (being able to generate descriptions of functionality anywhere on a given screen).} However, one aspect we observed across our models is that the most common observed types of functionality (\eg back buttons, menu buttons) corresponded to the functionalities that our models predicted most often and most confidently on unseen screenshots. This is somewhat expected, as the models saw the most examples of such functionalities during training. Thus, the diversity of predictions is an open problem for future research. This problem can be partially mitigated by larger, more diverse datasets with specifically curated descriptions (such as extensions to the \Clarity dataset). However, it is likely that domain-specific models, or ensembles of models, may be required to more effectively predict diverse features.

\textbf{\revision{Lesson 5: Future studies that evaluate automated GUI-centric documentation approaches should include human studies, as human perceptions of models may differ from automated reference-based metrics.}} One of the more surprising results of our study is that there seems to be a mismatch between humans perceptions of the captions generated by our DL models and the BLEU score metrics typically used to asses the accuracy of model predictions. This signifies that there are aspects of human perception that are not effectively captured in the BLEU metric, and possibly other translation metrics.

\vspace{-0.5em}
\revision{
\section{Limitations \& Threats to Validity}
\label{sec:limitations}
}
\revision{
\textbf{Internal Validity}. Threats to \textit{internal validity} correspond to unexpected factors in the experiments that may contribute to observed results. To derive our dataset we rely on MTurk and its workers to extract the high- and low-level descriptions per each screenshot. It should be noted that we did not ask MTurk workers to provide technical software documentation descriptions, but rather general descriptions of screen functionality at differing granularities. To minimize low quality captions we published the jobs for workers with more than 1k HITS, from English speaking countries, and HIT approval rate of more than 90\%. Also, each successfully completed HIT was vetted by at least one of the authors to assure quality. If there was any question related to caption quality, at least one of the other authors stepped in to resolve the ambiguity. As a result 2,429 HITs were rejected due to low quality descriptions.}

\revision{
\textbf{External Validity}. Threats to \textit{external validity} concern the generalization of the
results. As with any collected dataset, there is a threat to external validity about the generalizability of the \Clarity dataset.
However, we used a diverse set of popular apps from the Android domain, extracted popular screenshots from these apps, and the apps were captioned by a large and diverse set of MTurk workers. During our data collection process, we only collected 4 low-level captions per each screen in order to make the task feasible for MTurk workers as workers tend to abandon or perform poorly on long tasks. This means that, for certain screens with many GUI-components, some components may lack natural language descriptions. However, given the size of our dataset and the diversity of our screenshots and captions, we assert that our low-level captions are reasonably representative.}

\section{Related Work}
\label{sec:related-work}
\vspace{-0.0cm}

\noindent\textbf{DL for Image Captioning and GUIs}. Hossain \etal~\cite{Hossain:CSD'19} recently performed a wide-ranging study on DL models for image captioning, surveying the many different architectures and datasets used to evaluate them. However, this survey did not examine the ability of any image captioning model to predict functional descriptions of software. There have been a limited number of papers in the SE community that have applied DL techniques to GUI related data.  Chen \etal~\cite{Chen:ICSE'19} designed an approach that uses an NMT to translate an Android screenshot into a GUI-skeleton. However, their technique is able to predict GUI structure given an image, not functional natural language descriptions. Recently, Zhang et. al.~\cite{Zhang:UIST'21} created a dataset of iOS image captions to train a model for captioning accessibility data. However, the authors do not make their dataset publicly available and target a different goal of accessibility data compared our goal of generating functional captions. Chen~\etal investigated the use of DL image captioning models for applying labels to GUI-components in mobile apps~\cite{Chen:ICSE'20}, however, this approach only aims to predict short labels for a limited subset of GUI-components, whereas our study focuses upon predicting functional descriptions consisting of complete sentences for both individual GUI-components and entire screenshots.

\noindent\textbf{GUI-based Analysis of Mobile Apps}. 
GVT and GCat analyze the visual properties of GUIs to detect design violations and evolutionary changes~\cite{Moran:ASE'18,Moran:ICSE'18}. In contrast, we focus solely on image captioning techniques to provide functional program descriptions of screenshots. Approaches such as REMAUI~\cite{Nguyen:ASE'15b}, \ReDraw~\cite{Moran:TSE'18}, and \texttt{pix2code}~\cite{Beltramelli:arXiv17} aim to automatically generate mobile app code given an app screenshot. Conversely, we leverage DL techniques to generate functional descriptions rather than source code using a pixel-based image as input. Chen \etal~\cite{Chen:ASE'18} introduced StoryDroid, for automatically generating visual storyboards of Android apps to help aid in the app design process. However, their approach is not capable of generating a functional description of an application from GUI data.
Furthermore, Deka \etal showed how the Rico dataset could be navigated via semantic search using autoencoders ~\cite{Deka:UIST'17}. UiRef \cite{Andow:WiSec'17} is an approach for resolving security and privacy concerns by considering semantics of GUI-components that request user's inputs. Moreover, Liu \etal~\cite{Liu:UIST'18} presented an approach for automatically classifying mobile app icons according to semantic GUI patterns. Xiao \etal proposed IconIntent that combines program analysis and icon classification to detect privacy sensitive GUI-components \cite{Xiao:ICSE'19}. Different from this body of work, we aim to predict functional descriptions of GUIs for software documentation.

\vspace{-0.2cm}
\section{Conclusion}
\label{sec:conclusion}
\vspace{-0.2cm}

	In this paper, we have conducted one of the first comprehensive empirical investigations into the connection between GUI-related information, and functional descriptions of programs. We have derived the \Clarity dataset of GUI screenshots/metadata and NL captions, trained DL models on this dataset, and demonstrated their ability to bridge the semantic gap between visual and lexical program information.

\vspace{-0.5em}
\section*{Acknowledgment}
\vspace{-0.5em}
This work was supported by the NSF CCF-2007246 \& CCF-1955853 grants. Any opinions, findings, and conclusions expressed herein are the authors' and do not necessarily reflect those of the sponsors.

\balance
\bibliographystyle{ieeetr}
\bibliography{references}

\end{document}